\documentclass[aps,prb,manuscript]{revtex4}
\usepackage{graphicx}
\usepackage{epsfig}
\usepackage{color}
\usepackage{dcolumn}
\usepackage{bm}
\usepackage{amsmath}
\usepackage{amssymb}

\begin{document}
\draft

\title{Optical Modulation Effects on Nonlinear Electron Transport in Graphene in Terahertz Frequency Range}

\author{Danhong Huang,$^{1}$ Godfrey Gumbs,$^{2}$ and O. Roslyak$^{2}$}

\address{$^1$Air Force Research Laboratory, Space Vehicles Directorate, Kirtland Air Force Base, NM 87117, USA}

\address{$^2$Department of Physics and Astronomy, Hunter College at the City University of New York,
695 Park Avenue, New York, NY 10065, USA}

\date{\today}

\begin{abstract}
We describe very fast electron dynamics for a graphene nanoribbon
driven by a control electromagnetic field in the terahertz frequency
regime. The mobility  as a function of bias field
has been found to possess a large threshold value when entering a nonlinear
transport regime. This value depends on the lattice temperature, electron
density, impurity scattering strength, nanoribbon width and correlation
length for the line-edge roughness. An enhanced electron mobility beyond
this threshold has been observed, which is related to the initially-heated
electrons in high energy states with a larger group velocity. However, this
mobility enhancement quickly reaches a maximum governed by the Fermi velocity
in graphene and the dramatically increased phonon scattering. Super-linear
and sub-linear temperature dependences of the mobility are seen in the linear
and nonlinear transport regimes, which is attributed separately to the
results of sweeping electrons from the right Fermi edge to the left one
through elastic scattering and moving electrons from low-energy states
to high-energy ones through field-induced electron heating. The threshold
field is pushed up by a decreased correlation length in the high field regime,
and is further accompanied by a reduced magnitude in the mobility enhancement.
This implies an anomalous high-field increase of the line-edge
roughness scattering with decreasing correlation length due to the
occupation of high-energy states by field-induced electron heating. Additionally, a
self-consistent device modeling has been proposed for graphene transistors
under an optical modulation on its gate, which employs Boltzmann moment
equations up to the third order for describing fast carrier dynamics and
full wave electromagnetics coupled to the Boltzmann equation for describing spatial-temporal dependence of the total field.
Finally, a detailed comparison of the derived Maxwell-Boltzmann moment equations in this paper
with the well known Vlasov-Maxwell equations is also included.
\end{abstract}

\maketitle

\section{Introduction}
\label{sec1}

The engineering achievement of isolating graphene
 sheets\,\cite{art10,art11,art12,philtrans}
from graphite has inspired many studies aimed at understanding basic underlying physics\,\cite{art10,neto} as well as finding possible  applications to
carbon-based electronics\,\cite{art13}. The low-field linear transport
of charge carriers in a graphene layer, has received a considerable
amount of attention.\,\cite{art1,art2,art3,art4,art5,art7,art8} Recent reports on the
successful fabrication of ultra-fast graphene transistors\,\cite{art14} and
photodetector\,\cite{art15} has further advanced this research frontier into the
fields of electronics and optoelectronics. The graphene transistor was
reported to be used as both an electrical modulator\,\cite{modulator}
with a frequency as high as $\sim 10$\,GHz and as a sensitive
photo-detector for imaging\,\cite{detector}. However, similar investigations
 of linear transport in graphene nanoribbons (GNRs) have only been given relatively
little attention so far.\,\cite{art6,fang,art9}
\medskip

Early theoretical studies\,\cite{art7,fang} on electron transport in graphene
nanoribbons were restricted to the low-field regime, where a linearized Boltzmann
equation was solved within a relaxation-time approximation. In this paper,
the non-equilibrium distribution of electrons is calculated exactly by
solving the Boltzmann transport equation beyond the relaxation-time approximation
for nonlinear electron transport in semiconducting graphene nanoribbons.
Enhanced electron mobility from initially-heated electrons in high energy states
is anticipated. An anomalous increase in the line-edge roughness scattering for
large electric fields is obtained with decreasing roughness correlation length due
to the occupation of high-energy states by field-induced electron heating.
The semi-classical Boltzmann transport equation is expected to be applicable
to the diffusive band-transport regime with relative smooth edges for graphene
nanoribbons,  instead of the hopping and tunneling between localized states
with rough edges.
\medskip

Macroscopic simulation of semiconductor device physics has proceeded
with solving coupled Maxwellï-Boltzmann equations.\,\cite{matt-1,matt-2}
By employing the quasi-equilibrium Fermi-Dirac distribution in the
Boltzmann transport equation, we can obtain the  spatial dependence of
both chemical potential and temperature, which paves the ground for
drift-diffusion and hydrodynamic charge transport theories.\,\cite{matt-2,matt-3}
An ensemble of interacting electrons can be thermalized quickly with a
high density, which justifies the assumption of a quasi-equilibrium Fermi-Dirac
distribution for hot carriers. To determine the spatial dependence of chemical
 potential and temperature, device simulators usually couple Maxwell
equations for the electromagnetic fields to conservation relations
for carrier density, the current and energy.
Several investigations on hydrodynamics are related to determining the proper
mathematical representation of carrier thermal conductivity and the moments
of the Boltzmann equation.\,\cite{matt-2,matt-12,matt}
\medskip

The outline of the remainder of this paper is as follows. In Sec.\ \ref{SEC:Model},
we solve exactly  the semi-classical Boltzmann transport equation for
low-temperature electron transport in semiconducting graphene nanoribbons by including impurity, line-edge roughness and phonon scattering effects at a microscopic
level. Based on the calculated non-equilibrium distribution as a function of
wave number along the ribbon, we present detailed numerical results
for the electron mobility as a function of either the applied electric field
or the lattice temperature for various impurity scattering strengths and
correlation lengths for line-edge roughness. Our numerical results are
presented in Sec.\ \ref{sec3} with some discussions. In Sec.\ \ref{SEC:Model-1},
we derive the moment equations from the Boltzmann equation up to the third order
for electron dynamics in $n-$doped graphene as a generalization to
hydrodynamic model. At the same time, the self-consistent field
equations are also derived in Sec.\ \ref{SEC:Model-2} within the
Maxwell-Boltzmann frame. Finally, the conclusions of this paper are presented
in Sec.\ \ref{sec4}.

\section{Nonlinear Transport in Graphene Nanoribbons}

In this section, we  employ the Boltzmann transport model with inclusion
of scattering at a microscopic level  to study high-field nonlinear transport
of electrons in graphene nanoribbons along with some numerical results.

\subsection{Nonlinear Boltzmann Transport Model}
\label{SEC:Model}

Here, we  investigate single subband nonlinear transport only
in the armchair nanoribbon (ANR) configuration\,\cite{wakabayashi}. Low electron densities,
moderate temperatures, ionized impurities and line-edge roughness are considered\,\cite{fang,huang1,huang2}. As a result, negligible pair
scattering\,\cite{pair}, optical and out-of-plane flexural phonons\,\cite{huang2}, inter-valley scattering, volume-distributed and short-range impurity
scattering\,\cite{fang} will  all be neglected. Therefore, the electron-like
branch for $n-$doped graphene can be represented on a $k-$space mesh,\,\cite{neto}
via its dispersion and corresponding wave-function, as

\begin{gather}
\label{EQ:DISPERSION}
\varepsilon_j = \hbar\nu_F\,
\sqrt{k^2_j +\left({\pi/3W}\right)^2}\ ,\\
\label{EQ:WAVEFUNCTION}
\psi_{j}(x,\,y)=\sqrt{\frac{1}{2LW}}\,e^{ik_jy}\,
\left[{\begin{array}{c} 1\\ e^{i\phi_{j}}\end{array}}\right]\, \texttt{e}^{i(2\pi/3a_0-\kappa)x}\ .
\end{gather}
Here, $\nu_F=10^6$\,m/s is the Fermi velocity in graphene and
$L$ is the quantization length of the ribbon. For semiconducting ANR,
$\kappa=\pi/3W\ll 2\pi/3a_0$ is the quanta of the transverse wave
vector and $\phi_j=\tan^{-1}\left({k_j/\kappa}\right)$ is the phase
separation between the two graphene sublattices. The electron wave numbers
$k_j=\left[{j-(N+1)/2}\right]\,\delta k$ are given on the discrete mesh by
$j=1,\,2,\,\ldots,\,N$ for large odd integer $N$, $\delta k=2\,k_{\rm max}/(N-1)$
is a small mesh spacing, and $k_{\rm max}$ is chosen to ensure that
scattering induced population of higher electron-like branches can be neglected.
The minimum in the energy dispersion curve corresponds to the central mesh point $j=M=(N+1)/2$. Also, $W=(\mathcal{N}+1)\,a_0/2$ is the width of
a ribbon expressed in units of the size of graphene unit cell
$a_0=2.6$\,\AA\ and the number of carbon atoms $\mathcal{N}$ across the
ribbon. According to the dispersion relation in Eq.\,(\ref{EQ:DISPERSION}),
the electron group velocity $v_j$ for semiconducting ANRs is
given by  $v_j=\nu_F\left({\hbar\nu_{F}k_j/\varepsilon_j}\right)$.
Additionally, we assume that the electron-like branch is filled up to
$\lvert{k_j}\rvert=k_F$ at zero temperature with the Fermi wave number
and energy given by $k_F=\pi n_{1D}/2$ and $\varepsilon_F=\varepsilon(k_F)$,
respectively. For a chosen temperature $T$ and chemical potential
$\mu_0$, the linear electron density in ANR follows from $n_{1D}=\delta k/\pi\,\sum\limits_{j=1}^N\,f^{(0)}_j$, with $f^{(0)}_j = \left\{1+\exp\left[(\varepsilon_j-\mu_0)/\mathrm{k}_B T\right]\right\}^{-1}$
being the equilibrium Fermi-Dirac distribution function.
\medskip

Conventionally, the non-equilibrium carrier distribution function is
partitioned as $f_j=f^{(0)}_j+g_j$. The deviation from the equilibrium
Fermi distribution under a strong electric field
is described by the set of reduced nonlinear Boltzmann transport equations\,\cite{huang1,huang2}

\begin{equation}
\frac{dg^\prime_j(t)}{dt}=e v_j\,\mathcal{F}_0\,
\left(\frac{\partial f^{(0)}_j}{\partial \varepsilon_j}\right)
-\sum_{j^{\,\prime}\neq M}\,\mathcal{S}^\prime_{j,\,j^{\,\prime}}(t)\,g^\prime_{j^{\,\prime}}(t)\ .
\label{e1}
\end{equation}
In this notation, $g^\prime_j(t) = g_j(t)-g_M (t)$ is the reduced form
of the dynamical non-equilibrium part\,\cite{fn2} of the electron distribution
function. The reduced form accounts for particle number conservation condition,
i.e.,  $\sum \limits_{j=1}^N g_j(t) =0$. Reduced scattering matrix elements
$\mathcal{S}^\prime_{j,\,j^{\,\prime}}(t)=
\mathcal{S}_{j,\,j^{\,\prime}}(t)-\mathcal{S}_{j,\,M}(t)$
are defined via its components

\begin{eqnarray}
\mathcal{S}_{j,\,j^{\,\prime}}(t)&=&\delta_{j,\,j^{\,\prime}}\left[{\cal S}^{\texttt{in}}_{j,j'}+\bar{\cal S}^{\texttt{in}}_{j,j'}\{g'_{j'}\}+
{\cal S}^{\texttt{el}}_{j}\left({1-\delta_{j,\,(N+1)/2}}\right)\right]-
\delta_{j+j^{\,\prime},\,N+1}\left[ {\cal S}^{\texttt{el}}_{j} \left({1-\delta_{j,\,(N+1)/2}}\right)\right]
\nonumber\\
&-& {\cal S}^{\texttt{in}}_{j,\,j^{\,\prime}}-\frac{e{\cal F}_0}{2 \hbar\delta
k} \left({\delta_{j,j^\prime-1}-\delta_{j,j^\prime+1}}\right) \ .
\label{e7}
\end{eqnarray}
Here, the elastic scattering rate is given by:
\begin{gather}
\label{EQ:ELASTICSCATTERINRATE}
 {\cal S}^{\texttt{el}}_{j} = {\cal S}^{\texttt{imp}}_{j}+ {\cal S}^{\texttt{LER}}_{j}
= \left({\gamma_0 + \frac{\gamma_1}{1+4k_j^2\Lambda_0^2}}\right) \left(\frac{v_F}{|v_j|}\right)\,\left[1+\cos(2\phi_j)\right]\ ,
\end{gather}
where $\gamma_0 \sim n_{2D}$ denotes the impurity scattering rate
at the Fermi edges and $v_F=v_j(k_F)$. Since the momentum difference
between two valleys is rather large, only short-range impurities
(such as topological defects) with a range smaller than the lattice
constant will give rise to inter-valley scattering.
$\gamma_1 = 2\left(\pi\nu_F\delta b/3W^2\right)^2\left(\Lambda_0/\nu_F\right)$
is the scattering rate due to edge roughens, with $\delta b\sim 5$\,\AA\ being the amplitude and $\Lambda_0$ being the correlation length of the roughness.
\medskip

The dominating inelastic scattering mechanism is provided by longitudinal
acoustic phonons at low temperatures. The static inelastic scattering rates
are given by the following matrix elements

 \begin{eqnarray}
\label{e14}
{\cal S}^{\texttt{in}}_{j,\,j^{\,\prime}} & = & \frac{L}{2\pi}\,\delta k \sum_\pm\,{\cal S}^\pm_{j,j^\prime}\,\left({n_{j,j^\prime}+f^{\pm}_{j^\prime}}\right)\ ,\\
S^{\pm}_{j,j^\prime} & = &
\theta(\pm\varepsilon_{j^{\,\prime}}\mp\varepsilon_j)\,\left[
\frac{D_{AL}^2|\varepsilon_{j^{\,\prime}}-\varepsilon_j|}{2\hbar^2c_s^3
\rho_ML W \epsilon^2_{\rm TF}(|k_{j^{\,\prime}}-k_j|)}\right]\,
\left[1+\cos(\phi_{j^{\,\prime}}-\phi_j)\right]\ .
\end{eqnarray}
Here, $f^{-}_j=f^{(0)}_j$, $f^{+}_j=1-f^{(0)}_j$, $n_{j,j^\prime}
=N_0(|\varepsilon_{j^{\,\prime}}-\varepsilon_j|/
\hbar)$, $N_0(\omega_q)=[\exp(\hbar\omega_q/k_{\rm B}T)-1]^{-1}$ is the
Bose-Einstein function for thermal equilibrium phonons;
$D_{AL}\sim 16$\,eV is the deformation potential,
$\rho_M\sim 7.6\times 10^{-8}$\,g/cm$^2$ and $c_s\sim 2\times 10^6$\,cm/s
are the mass density and sound velocity in graphene.
The scattering potentials are screened
by free carriers and described by a dielectric function. Here, we assume
that the inelastic scattering is shielded by the static Thomas-Fermi
dielectric function in its general form\,\cite{fang,fertig1}
$\epsilon_{TF}(\lvert{k_{j^\prime}-k_j}\rvert)$. The screening of elastic
scattering potentials is given approximately by $\epsilon_{TF}\approx 1+(e^2/\pi^2\epsilon_0\epsilon_r\hbar\nu_F)$ under the metallic limit
($2k_F W\gg 1$) with $\epsilon_r\approx 3.9$.
\medskip

The nonlinear dynamical phonon scattering rate  is

\begin{eqnarray}
\bar{\cal S}^{\texttt{in}}_{j,j'}\{g'_{j'}\}=\frac{L}{2\pi}\,\delta k\,
g^\prime_{j^{\,\prime}}(t) \left[{\cal S}^{+}_{j,j^\prime}
-{\cal S}^{-}_{j,j^\prime}-\left({\cal S}^{+}_{j,M}
-{\cal S}^{-}_{j,M}\right)\right]\ ,
\label{e17}
\end{eqnarray}
which is also responsible for the nonlinear electron transport and electron
heating due to its dependence on $g^\prime_{j^{\,\prime}}(t)$.
\medskip

Once the non-equilibrium part, $g^\prime_j(t)$, of the total electron
distribution function has been determined  using Eq.\,(\ref{e1}), the
transient drift velocity, $v_c(t)$, of the system can be calculated
with the use of

\begin{equation}
\label{EQ:DRIFTVELOCITY}
v_{\rm c}(t)=\left[{\sum\limits_{j=1}^{N}\,f^{(0)}_j}\right]^{-1}\,
\sum\limits_{j\neq M}\,\left(v_j -v_M\right)\,g^\prime_j(t)\ .
\end{equation}
We note that the thermal-equilibrium part of the electron distribution
does not contribute to the drift velocity. The steady-state drift velocity
$v_{\rm d}$ of electrons is given by $v_{\rm c}(t)$ by taking
 the limit $t \to \infty$.  The corresponding steady-state conduction
 current is given by  $I=en_{1D}v_{ \rm d}$. The differential electron
 mobility for  nonlinear transport is generalized to
$\mu_e=\partial v_{\rm d}/\partial\mathcal{F}_0$.
Numerical simulation of these quantities in specific ANRs is presented below.

\subsection{Numerical Results and Discussion}
\label{sec3}

\begin{figure}[p]
\epsfig{file=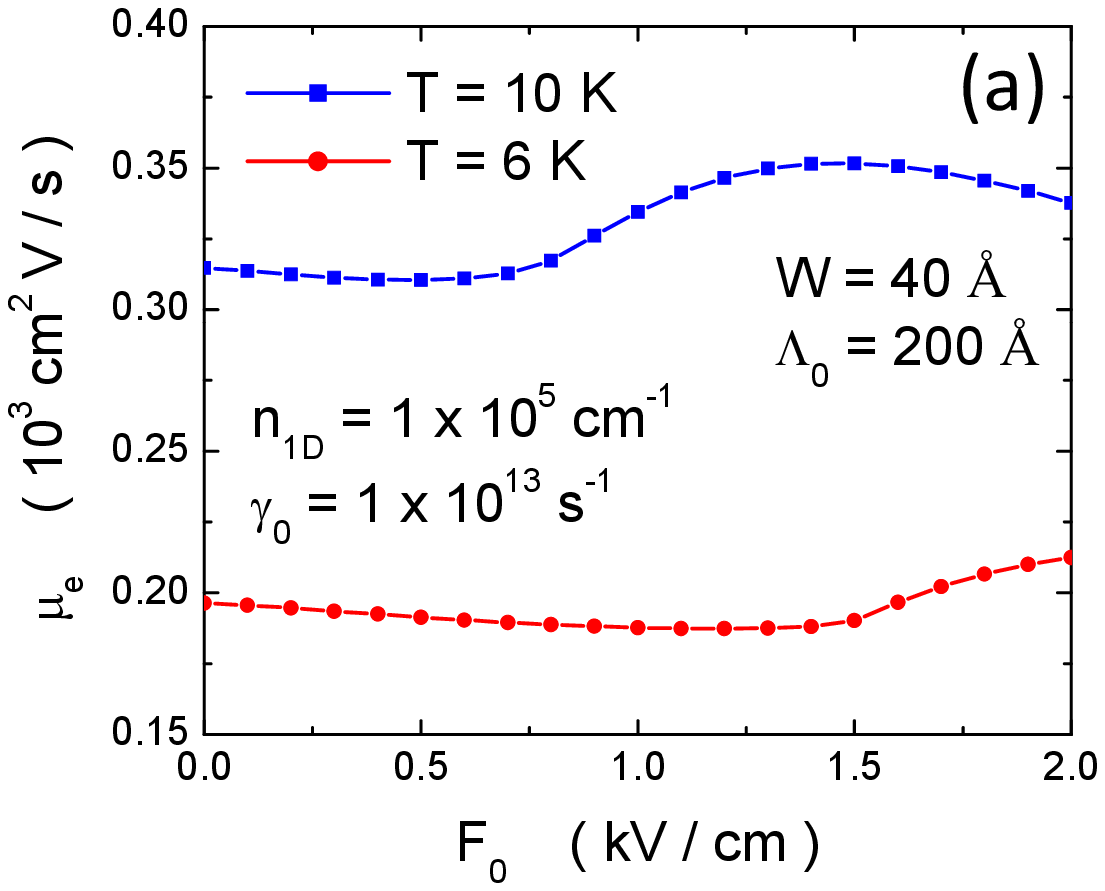,width=8cm}
\epsfig{file=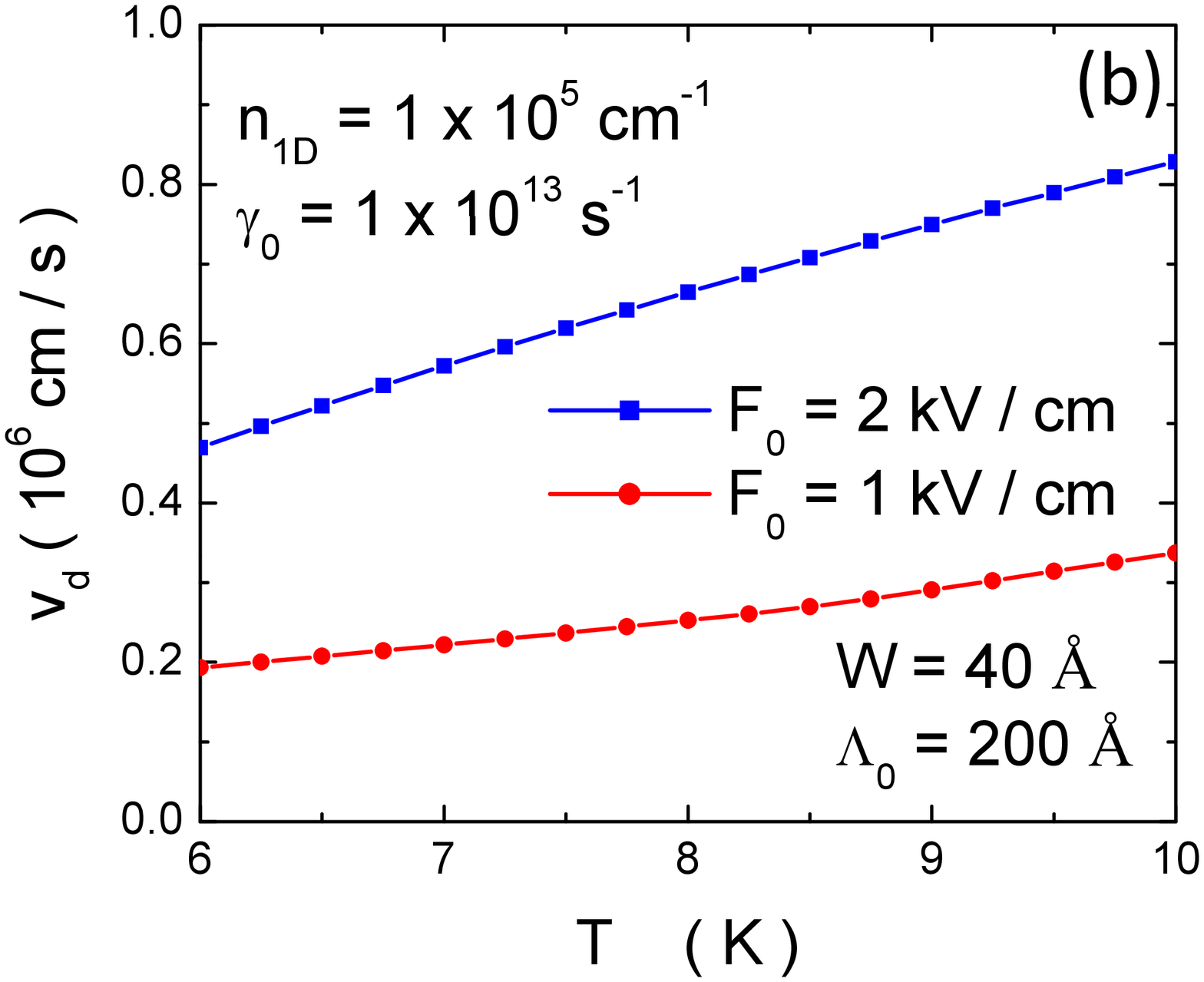,width=8cm}
\caption{(Color online) (a) Calculated electron mobilities $\mu_{\rm e}$ as a function of applied
electric field ${\cal F}_0$ at $T=10$\,K (solid squares on blue curve) and $T=6$\,K
(solid circles on red curve); (b) Electron drift velocities $v_{\rm d}$ as a function of
temperature  $T$ at ${\cal F}_0=2$\,kV/cm (solid squares on blue curve) and ${\cal F}_0=1$\,kV/cm (solid circles on red curve).}
\label{f1}
\end{figure}

Figure\ \ref{f1}(a) presents our  calculated electron mobilities $\mu_{\rm e}$ as
a function of applied electric field ${\cal F}_0$ at $T=10$\,K (blue curve)
and $T=6$\,K (red curve), respectively. Clearly, from Fig.\,\ref{f1}(a),  a strong
${\cal F}_0$-dependence for $\mu_{\rm e}$ appears at a lower value of ${\cal F}_0$
at higher temperature $T$ in the nanoribbon. This ${\cal F}_0$-dependent electron
mobility $\mu_{\rm e}$ has its physical origins  in the dynamical
electron-phonon scattering rate $\bar{\cal
S}^{\texttt{in}}_{j,j'}\{g'_{j'}\}$ through its dependence on the electron
distribution  function given in Eq.\,(\ref{e17}). Consequently, it is reasonable to
expect a lower threshold field, ${\cal F}^\ast$, for entering into a nonlinear
transport regime (${\cal F}>{\cal F}^\ast$) due to enhanced nonlinear phonon
scattering at $T=10$\,K.  The value of ${\cal F}^\ast$ strongly depends on the
parameters of the system, such as $T$, $n_{\rm 1D}$, $\gamma_0$ and $\Lambda_0$, and
an analytic expression for ${\cal F}^\ast$ cannot be obtained in the
nonlinear transport regime. On the other hand, as ${\cal F}_0\to 0$, $\mu_{\rm  e}$
is larger at $T=10$\,K than  $T=6$\,K due to thermal population of
high-energy states with a large electron group velocity. The initial decrease of
$\mu_{\rm e}$ with ${\cal F}_0$ is attributed to the gradual increase of
the frictional force from phonon scattering by ${\cal F}_0$. At
$T=10$\,K, $\mu_{\rm e}$ is roughly independent of
${\cal F}_0$ below $0.75$\,kV/cm (linear regime). However, $\mu_{\rm e}$ increases
significantly with ${\cal F}_0$ above $0.75$\,kV/cm (nonlinear regime). Eventually,
$\mu_{\rm e}$ decreases with ${\cal F}_0$ once it exceeds $1.5$\,kV/cm
(heating regime), leading to a saturation of the electron drift velocity.
The electron group velocity $|v_j|$ increases with the wave number for small $|k_j|$ values, as can be seen from Eq.\,(\ref{EQ:DISPERSION}).
However, the increase of $|v_j|$ slows down toward its upper limit
$\nu_F$ provided $|k_j|\gg\pi/3W$ but still within the single-subband regime.
The increase of $\mu_{\rm e}$ with ${\cal F}_0$ in the nonlinear regime comes from the
initially-heated electrons in high energy states with a larger group velocity,
while the successive decrease of $\mu_{\rm e}$ in the heating regime comes from
the combination of the dramatically increased phonon scattering
and the upper limit $\nu_F$ mentioned above. In Fig.\,\ref{f1}(b), the
calculated electron drift velocities $v_{\rm d}$ are plotted as a functions
of temperature when ${\cal F}_0=2$\,kV/cm (blue curve) and
${\cal F}_0=1$\,kV/cm (red curve). The fact that $\mu_{\rm e}$ increases with $T$ monotonically in both cases implies the electron scattering in
two samples is not dominated  by phonons but by impurities and line-edge roughness.
Different behaviors in the increase of $\mu_{\rm e}$ with temperature can be
seen from Fig.\,\ref{f1}(b) for linear and nonlinear electron transport. At ${\cal F}_0=2$\,kV/cm for the high-field nonlinear transport, $v_{\rm d}$ (or $\mu_{\rm e}$) increases with $T$ sub-linearly. On the other hand, $v_{\rm d}$ rises
super-linearly with $T$ for the low-field linear transport at
${\cal F}_0=1$\,kV/cm. These different $T$ dependence in $\mu_{\rm e}$
for linear and  nonlinear transports can be directly related to the
 non-equilibrium part of the electron distribution  function $g_j$.
\medskip

\begin{figure}[p]
\begin{center}
\epsfig{file=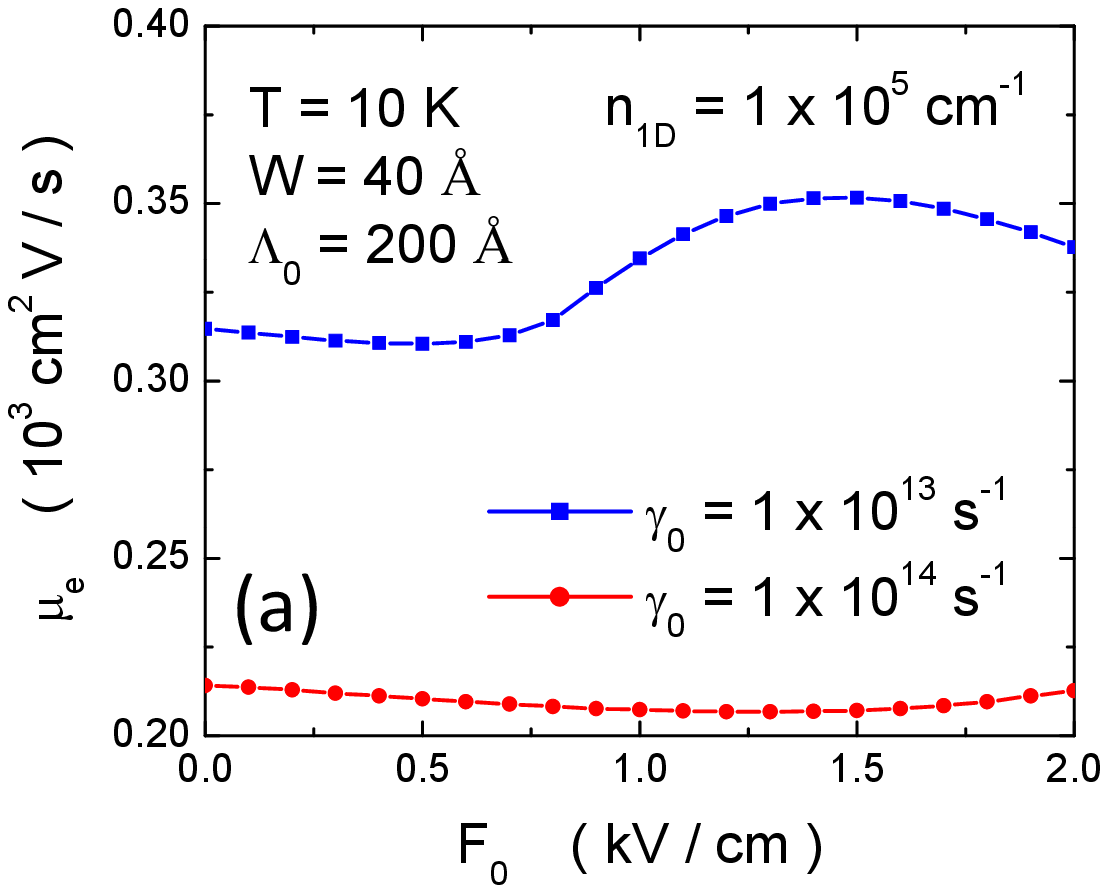,width=8cm}
\epsfig{file=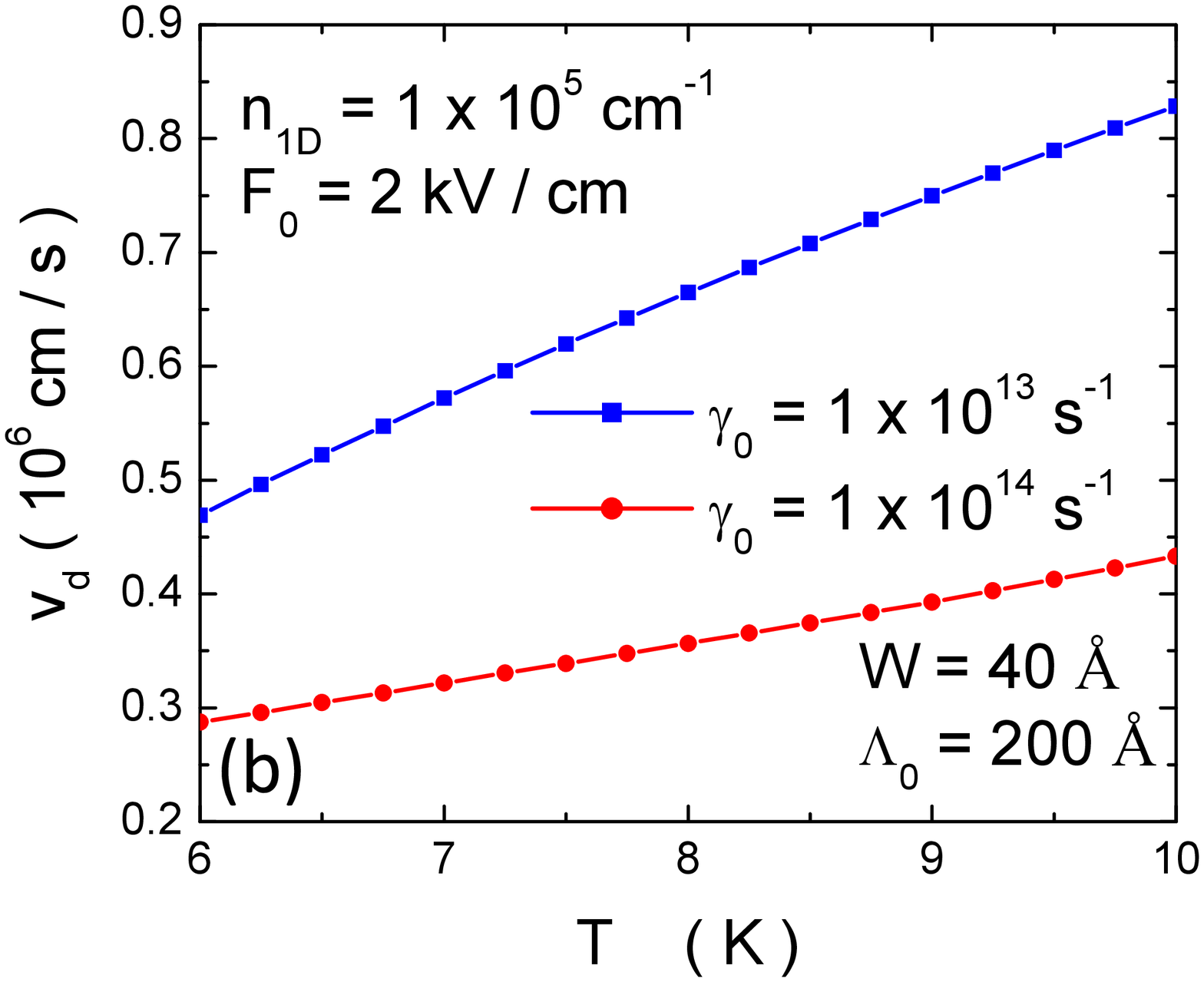,width=8cm}
\end{center}
\caption{(Color online) (a) $\mu_{\rm e}$ as a function of ${\cal F}_0$ at $T=10$\,K with
$\gamma_0=1.0\times 10^{13}$\,s$^{-1}$ (solid squares on blue curve)
and $\gamma_0=1.0\times 10^{14}$\,s$^{-1}$ (solid circles on red curve);
(b) $v_{\rm d}$ as a function of $T$ with ${\cal F}_0=2$\,kV/cm for
$\gamma_0=1.0\times 10^{13}$\,s$^{-1}$ (solid squares on blue curve)
and $\gamma_0=1.0\times 10^{14}$\,s$^{-1}$ (solid circles on red curve).}
\label{f3}
\end{figure}

The effects due to impurity scattering are compared in Figs.\,\ref{f3}(a)
and \ref{f3}(b). In Fig.\,\ref{f3}(a), we present a comparison of mobilities
as a function of ${\cal F}_0$ at $T=10$\,K for
$\gamma_0=1.0\times 10^{13}$\,s$^{-1}$ (blue curve) and
$\gamma_0=1.0\times 10^{14}$\,s$^{-1}$ (red curve). As
${\cal F}_0\to 0$, $\mu_{\rm e}$ is  greatly reduced by strong
impurity scattering with $\gamma_0=1.0\times 10^{14}$\,s$^{-1}$ in
the linear regime. In addition, for $\gamma_0=1.0\times 10^{14}$\,s$^{-1}$,
${\cal F}^\ast$ is pushed upward from about $0.75$\,kV/cm to $1.75$\,kV/cm,
leaving us with a roughly ${\cal F}_0$-independent $\mu_{\rm e}$ in this case
for the whole field range shown in this figure. The comparison for $T$-dependence
of $\mu_{\rm e}$ is presented in Fig.\,\ref{f3}(b) for ${\cal F}_0=2$\,kV/cm, where a sub-linear increase of $\mu_{\rm e}$ with $T$ for weak impurity scattering is
switched to a super-linear relation in the strong impurity scattering case.
\medskip

\begin{figure}[p]
\begin{center}
\epsfig{file=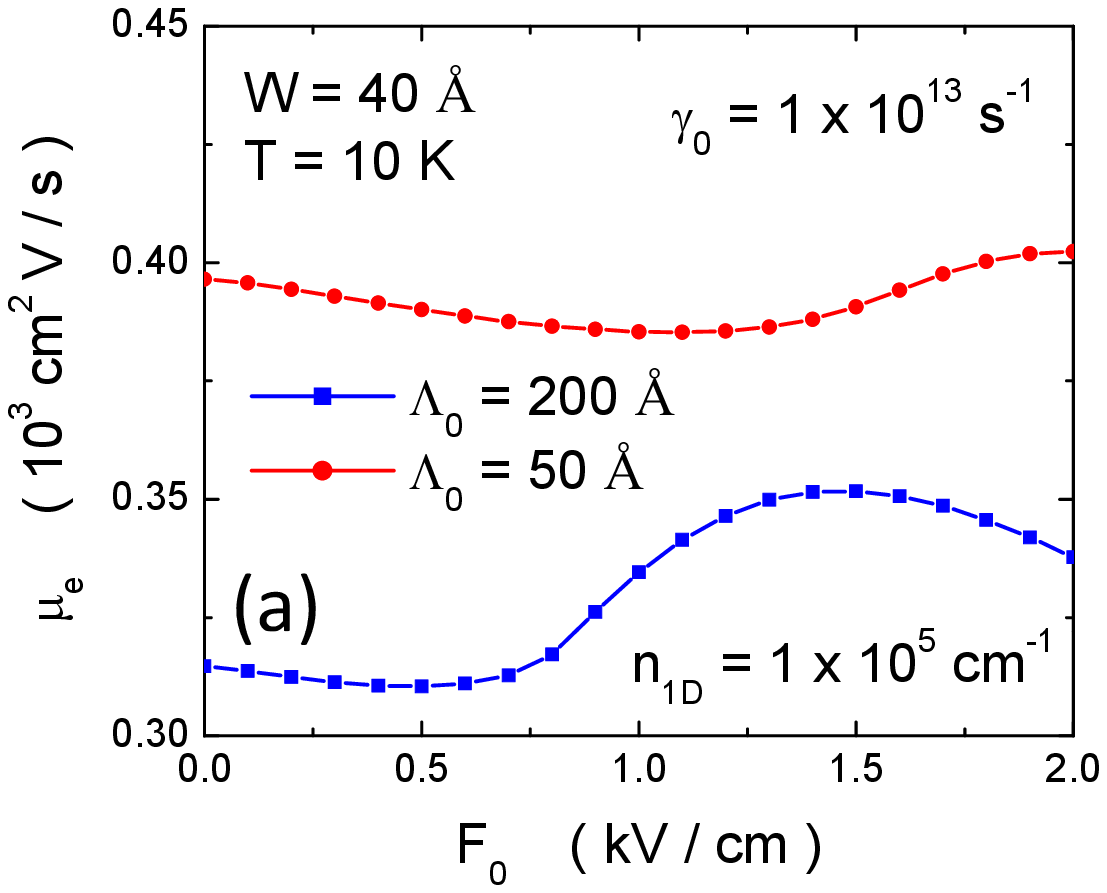,width=8cm}
\epsfig{file=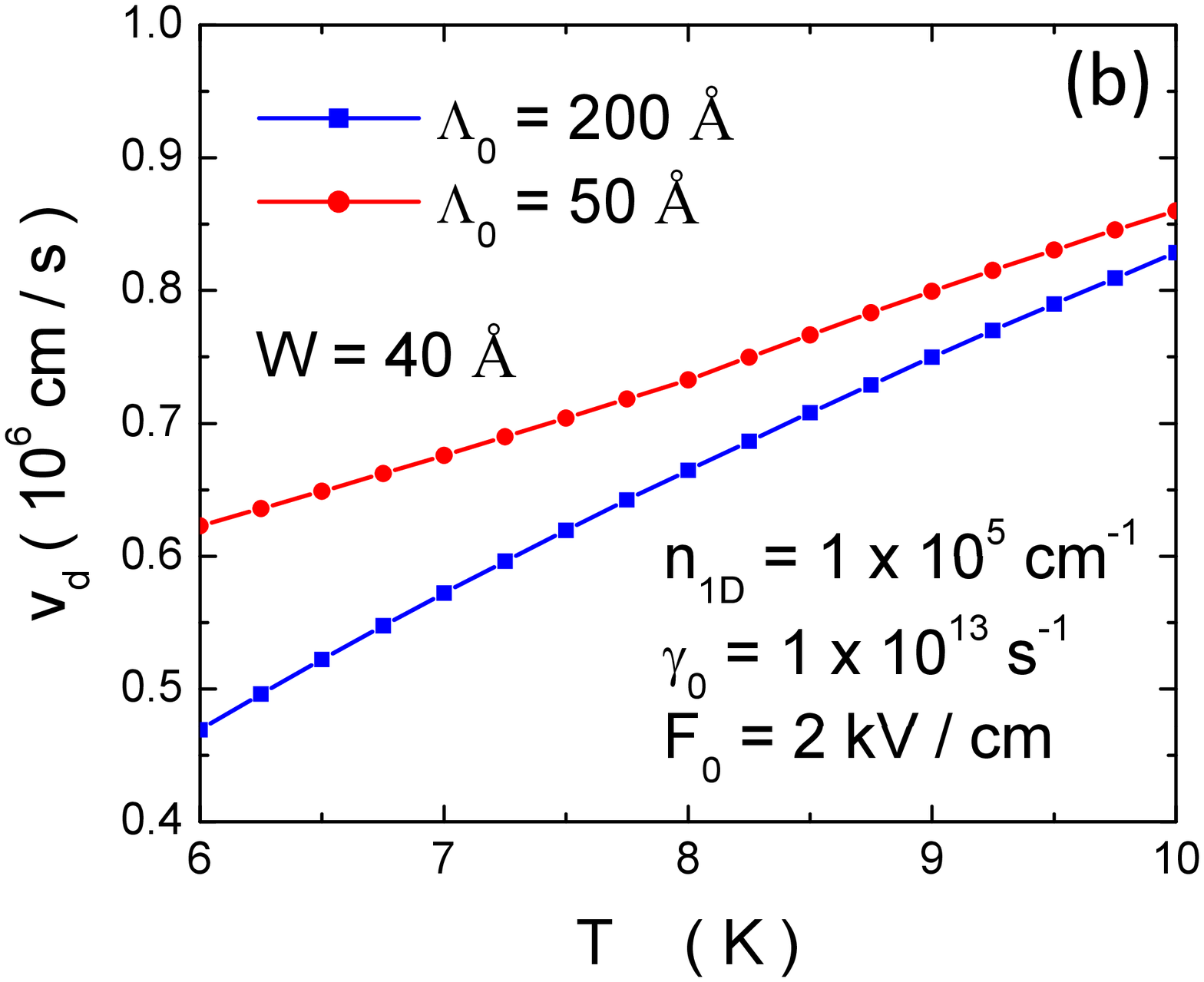,width=8cm}
\end{center}
\caption{(Color online) (a) $\mu_{\rm e}$ as a function of ${\cal F}_0$ at $T=10$\,K with
$\Lambda_0=200$\,\AA\ (solid squares on blue curve) and $\Lambda_0=50$\,\AA\ (solid circles on
red curve); (b) $v_{\rm d}$ as a function of $T$ with ${\cal F}_0=2$\,kV/cm for
$\Lambda_0=200$\,\AA\ (solid squares on blue curve) and $\Lambda_0=50$\,\AA\ (solid circles on red curve).}
\label{f5}
\end{figure}

Finally, the effect of correlation length for the line-edge roughness on
the transport is demonstrated in Figs.\,\ref{f5}(a) and \ref{f5}(b) by
fixing $W=50$\,\AA\ and changing $\Lambda_0$ from $200$\,\AA\ to $50$\,\AA.
As shown in Eq.\,(\ref{EQ:ELASTICSCATTERINRATE}), the line-edge roughness
scattering can be either reduced or enhanced by decreasing $\Lambda_0$,
depending on $|k_j|\ll 1/2\Lambda_0$ or $|k_j|\gg 1/2\Lambda_0$.
For our chosen sample with $n_{\rm 1D}=1.0\times 10^5$\,cm$^{-1}$, we find
that the condition $|k_j|\ll 1/2\Lambda_0$ is satisfied in the low-field
limit ($|k_j|\sim k_{\rm F}$), while $|k_j|\gg 1/2\Lambda_0$  holds for
the  high field limit due to electron heating. Therefore, we find from
Fig.\,\ref{f5}(a) that $\mu_{\rm e}$ is increased as ${\cal F}_0\to 0$ when
$\Lambda_0$ is reduced to $50$\,\AA\ in the low field regime. However, the
 value of ${\cal F}^\ast$ for $\mu_{\rm e}$ is pushed upward for
 $\Lambda_0=50$\,\AA\ in the high-field regime, which is further accompanied by a
reduced magnitude in the enhancement of $\mu_{\rm e}$ with ${\cal F}_0$.
This anomalous feature associated with reducing $\Lambda_0$ also has a profound
impact on the $T$-dependence of $\mu_{\rm e}$ as shown in Fig.\,\ref{f5}(b),
where the increasing rate of $\mu_{\rm e}$ with $T$ in the high field regime
becomes much lower with $\Lambda_0=50$\,\AA\ than for $\Lambda_0=200$\,\AA.

\section{Optical Modulation to Graphene Layers}

In this section, we will first derive the Boltzmann moment equations up to
 third order in  describing very fast electron dynamics driven by a control
 electromagnetic field in the terahertz frequency regime. At the same time, a self-consistent equation is also derived for the total driven field, including
the induced optical coherence in a graphene layer.

\subsection{Boltzmann Moment Equations}
\label{SEC:Model-1}

For electrons in a two-dimensional $n-$doped conducting graphene layer
with its band structure given by a Dirac cone, the Boltzmann
equation for the electron distribution function $f({\bf r}_\|,\,{\bf
k}_\|,\,t)$ around ${\bf K}$ (or ${\bf K}^\prime$) valley point is

\begin{equation}
\frac{\partial f({\bf r}_\|,\,{\bf k}_\|,\,t)}{\partial
t}+\frac{d{\bf r}_\|}{dt}\cdot\nabla_{{\bf r}_\|}f({\bf r}_\|,\,{\bf
k}_\|,\,t)+\frac{d{\bf k}_\|}{dt}\cdot\nabla_{{\bf k}_\|}f({\bf r}_\|,\,{\bf
k}_\|,\,t)=\left.\frac{\partial f({\bf r}_\|,\,{\bf k}_\|,\,t)}{\partial
t}\right|_{\rm coll}\ , \label{d1}
\end{equation}
where ${\bf v}_{k_\|}=d{\bf r}_\|/dt=\nabla_{{\bf
k}_\|}\varepsilon_{k_\|}/\hbar=\nu_F\,({\bf k}_\|/k_\|)$ is the
group velocity of Bloch electrons with kinetic energy
$\varepsilon_{k_\|}=\hbar\nu_Fk_\|$ and the Fermi velocity
$\nu_F\sim 10^8$\,cm/s . The Newton's second law requires that
$\hbar d{\bf k}_\|/dt={\bf F}_{k_\|}=-e\,({\bf E}_\|+{\bf
v}_{k_\|}\times{\bf B}_0)$ with $-e$, ${\bf E}_\|$ and ${\bf B}_0$
being the electron charge, in-plane electrical and static magnetic
fields, respectively. ${\bf E}_\|({\bf r}_\|,\,t)$ refers to the electric component
of the total electromagnetic field, which should be determined by the self-consistent field equation
(see next subsection below). The Markovian collision term introduced in
Eq.\,(\ref{d1}) is

\begin{equation}
\left.\frac{\partial f({\bf r}_\|,\,{\bf k}_\|,\,t)}{\partial
t}\right|_{\rm coll}={\cal W}^{\rm (in)}_{{\bf k}_\|}\left[1-f({\bf
r}_\|,\,{\bf k}_\|,\,t)\right]-{\cal W}^{\rm (out)}_{{\bf k}_\|}\,f({\bf
r}_\|,\,{\bf k}_\|,\,t)\ , \label{d2}
\end{equation}
where ${\cal W}^{\rm (in)}_{{\bf k}_\|}$ and ${\cal W}^{\rm (out)}_{{\bf
k}_\|}$ are the scattering-in and scattering-out rates for electrons in
the two-dimensional ${\bf k}_\|$-state. Based on the Newton's second law, we
rewrite Eq.\,(\ref{d1}) into the form of

\begin{equation}
\frac{\partial f({\bf r}_\|,\,{\bf k}_\|,\,t)}{\partial
t}=-{\bf v}_{k_\|}\cdot\nabla_{{\bf r}_\|}f({\bf r}_\|,\,{\bf k}_\|,\,t)-\frac{{\bf
F}_{k_\|}}{\hbar}\cdot\nabla_{{\bf k}_\|}f({\bf r}_\|,\,{\bf
k}_\|,\,t)+\left.\frac{\partial f({\bf r}_\|,\,{\bf k}_\|,\,t)}{\partial
t}\right|_{\rm coll}\ . \label{d3}
\end{equation}
\medskip

The zeroth-order moment of the Boltzmann equation in
Eq.\,(\ref{d3}) is found from

\[
\frac{2}{{\cal A}}\sum\limits_{{\bf k}_\|}\,\frac{\partial f({\bf
r}_\|,\,{\bf k}_\|,\,t)}{\partial t}=-\nabla_{{\bf
r}_\|}\cdot\left[\frac{2}{{\cal A}}\sum\limits_{{\bf k}_\|}\,{\bf
v}_{k_\|}\,f({\bf r}_\|,\,{\bf
k}_\|,\,t)\right]-\frac{1}{\hbar}\left[\frac{2}{{\cal
A}}\sum\limits_{{\bf k}_\|}\,{\bf F}_{k_\|}\cdot\nabla_{{\bf k}_\|}f({\bf r}_\|,\,{\bf
k}_\|,\,t)\right]
\]
\begin{equation}
+\frac{2}{{\cal A}}\sum\limits_{{\bf k}_\|}\,{\cal W}^{\rm (in)}_{{\bf
k}_\|}\left[1-f({\bf r}_\|,\,{\bf k}_\|,\,t)\right]-\frac{2}{{\cal
A}}\sum\limits_{{\bf k}_\|}\,{\cal W}^{\rm (out)}_{{\bf k}_\|}\,f({\bf
r}_\|,\,{\bf k}_\|,\,t)\ , \label{d4}
\end{equation}
which leads to the following electron number conservation equation, after the inter-valley scattering
is ignored,

\begin{equation}
\frac{\partial\rho({\bf r}_\|,\,t)}{\partial
t}=-\nabla_{{\bf r}_\|}\cdot{\bf j}_\|({\bf r}_\|,\,t)\ ,
\label{d5}
\end{equation}
where ${\cal A}$ is the sample area, $\rho({\bf r}_\|,\,t)=(2/{\cal
A})\sum\limits_{{\bf k}_\|}f({\bf r}_\|,\,{\bf k}_\|,\,t)$ is the
electron sheet number density (per area) and ${\bf j}_\|({\bf
r}_\|,\,t)=(2/{\cal A})\sum\limits_{{\bf k}_\|}{\bf v}_{k_\|}f({\bf
r}_\|,\,{\bf k}_\|,\,t)$ is the electron surface number current
density (per length). Equation\ (\ref{d5}) allows us to determine the
the spatial distribution of $\rho({\bf r}_\|,\,t)$ at each time $t$.
\medskip

In order to simplify the first-order moment of the Boltzmann
equation, we introduce the momentum-relaxation time
approximation. Under this approximation, we write

\begin{equation}
\left.\frac{\partial f({\bf r}_\|,\,{\bf k}_\|,\,t)}{\partial
t}\right|_{\rm coll}=-\frac{f({\bf r}_\|,\,{\bf
k}_\|,\,t)-f_0(\varepsilon_{k_\|},\,T,\,\mu_0)}{\tau_1}\ , \label{d6}
\end{equation}
where
$f_0(\varepsilon_{k_\|},\,T,\,\mu_0)=\{\exp[(\varepsilon_{k_\|}-\mu_0)/k_{\rm
B}T]+1\}^{-1}$ is the Fermi-Dirac function for thermal-equilibrium
electrons and $\tau_1$ is the average momentum-relaxation time for electrons.
In principle, $\tau_1$ can be microscopically calculated based on
${\cal S}_{j,j'}(t)$ introduced in Eq.\,(\ref{e7}) in the previous section for fixed applied bias and temperature
as well as device parameters. In addition, we
introduce the force-balance equation, which yields

\begin{equation}
{\bf k}_\|=-\frac{e\,\tau_1}{\hbar}\left({\bf E}_\|+{\bf
v}_{k_\|}\times{\bf
B}_0\right)=\left(\frac{\tau_1}{\hbar}\right){\bf F}_{k_\|}\ .
\label{d7}
\end{equation}
This leads to, to the leading order of a weak ${\bf B}_0$ field with
$\nu_FB_0/E_\|\ll 1$,

\begin{equation}
{\bf k}_\|\approx-\frac{e\tau_1}{\hbar}\left[{\bf
E}_\|+\nu_F\left(\frac{{\bf E}_\|\times{\bf
B}_0}{E_\|}\right)\right]\ . \label{d8}
\end{equation}
Employing the result in Eq.\,(\ref{d6}), we arrive at the first-order
moment of the Boltzmann equation

\[
\frac{2}{{\cal A}}\sum\limits_{{\bf k}_\|}\,{\bf v}_{k_\|}\,f({\bf r}_\|,\,{\bf
k}_\|,\,t)+\tau_1\,\frac{\partial}{\partial t}\left[\frac{2}{{\cal
A}}\sum\limits_{{\bf k}_\|}\,{\bf v}_{k_\|}\,f({\bf r}_\|,\,{\bf
k}_\|,\,t)\right]
\]
\[
=-\tau_1\left[\frac{2}{{\cal
A}}\sum\limits_{{\bf k}_\|}\,{\bf v}_{k_\|}\left({\bf v}_{k_\|}\cdot\nabla_{{\bf r}_\|}\right)f({\bf r}_\|,\,{\bf
k}_\|,\,t)\right]-\frac{\tau_1}{\hbar}\left[\frac{2}{{\cal A}}\sum\limits_{{\bf
k}_\|}\,{\bf v}_{k_\|}\left({\bf F}_{k_\|}\cdot\nabla_{{\bf k}_\|}\right)f({\bf r}_\|,\,{\bf
k}_\|,\,t)\right]
\]
\begin{equation}
+\frac{2}{{\cal A}}\sum\limits_{{\bf k}_\|}\,{\bf
v}_{k_\|}\,f_0(\varepsilon_{k_\|},\,T,\,\mu_0)\ . \label{d9}
\end{equation}
Approximating $f({\bf r}_\|,\,{\bf k}_\|,\,t)$ on the right-hand-side of
Eq.\,(\ref{d9}) by $f_0(\varepsilon_{k_\|},\,T,\,\mu_0)$, we get

\[
{\bf j}_\|({\bf r}_\|,\,t)+\tau_1\,\frac{\partial{\bf j}_\|({\bf
r}_\|,\,t)}{\partial t}
\]
\[
=-\tau_1\left[\frac{2}{{\cal A}}\sum\limits_{{\bf k}_\|}{\bf v}_{k_\|}\left({\bf
v}_{k_\|}\cdot\nabla_{{\bf
r}_\|}\right)f_0(\varepsilon_{k_\|},\,T,\,\mu_0)\right]-\tau_1\left[\frac{2}{{\cal
A}}\sum\limits_{{\bf k}_\|}\,{\bf v}_{k_\|}\left({\bf F}_{k_\|}\cdot{\bf
v}_{k_\|}\right)\frac{\partial
f_0(\varepsilon_{k_\|},\,T,\,\mu_0)}{\partial\varepsilon_{k_\|}}\right]
\]
\[
=-\frac{\nu_F^2\tau_1}{2}\,\nabla_{{\bf
r}_\|}\left[\frac{2}{{\cal A}}\sum\limits_{{\bf
k}_\|}\,f_0(\varepsilon_{k_\|},\,T,\,\mu_0)\right]
\]
\begin{equation}
+\frac{\nu_F^2e\tau_1}{2}\left[{\bf E}_\|+\nu_F\left(\frac{{\bf
E}_\|\times{\bf B}_0}{E_\|}\right)\right]\left[\frac{2}{{\cal
A}}\sum\limits_{{\bf k}_\|}\,\frac{\partial
f_0(\varepsilon_{k_\|},\,T,\,\mu_0)}{\partial\varepsilon_{k_\|}}\right]\
. \label{d10}
\end{equation}
It is straight forward to show for each valley that

\begin{equation}
\frac{2}{{\cal A}}\sum\limits_{{\bf
k}_\|}\,f_0(\varepsilon_{k_\|},\,T,\,\mu_0)=N_c\left(k_{\rm
B}T\right)^2\,{\cal Q}_1(\eta)\approx\rho({\bf r}_\|,\,t)\ ,
\label{d11}
\end{equation}

\begin{equation}
\frac{2}{{\cal A}}\sum\limits_{{\bf
k}_\|}\,\frac{\partial
f_0(\varepsilon_{k_\|},\,T,\,\mu_0)}{\partial\varepsilon_{k_\|}}=-N_c\left(k_{\rm
B}T\right)\,{\cal Q}_0(\eta)\ , \label{d12}
\end{equation}
where $\eta=\mu_0/k_{\rm B}T$,
$N_c=1/(\pi\hbar^2\nu_F^2)$, and the dimensionless
function

\begin{equation}
{\cal
Q}_n(\eta)=\int\limits_0^\infty\,\frac{x^n\,dx}{e^{(x-\eta)}+1}\
. \label{d13}
\end{equation}
The local chemical potential $\mu_0({\bf r}_\|,\,t)$ can be calculated from
Eq.\,(\ref{d11}) if both $T({\bf r}_\|,\,t)$ and $\rho({\bf r}_\|,\,t)$ are
given. Based on the results in Eqs.\,(\ref{d11}) and (\ref{d12}), we
finally get the generalized drift-diffusion equation

\[
\frac{\partial {\bf j}_\|({\bf r}_\|,\,t)}{\partial
t}=-\frac{{\bf j}_\|({\bf r}_\|,\,t)}{\tau_1}
\]
\begin{equation}
-N_c\,\frac{\nu_F^2e}{2}\left\{\left[{\bf
E}_\|+\nu_F\left(\frac{{\bf E}_\|\times{\bf
B}_0}{E_\|}\right)\right]\left(k_{\rm B}T\right)\,{\cal
Q}_0(\eta)+\frac{1}{e}\,\nabla_{{\bf r}_\|}\left[\left(k_{\rm
B}T\right)^2\,{\cal Q}_1(\eta)\right]\right\}\ . \label{d14}
\end{equation}
Equation\ (\ref{d14}) enables us to determine the spatial
distribution of the electron number current density ${\bf j}_\|({\bf
r}_\|,\,t)$ at each time $t$. The Einstein relation can be obtained
by setting $\partial\rho({\bf r}_\|,\,t)/\partial t=0$ and
$\partial{\bf j}_\|({\bf r}_\|,\,t)/\partial t=0$ in
Eqs.\,(\ref{d5}) and (\ref{d14}) for a steady state, i.e.
$\nabla_{{\bf r}_\|}\cdot{\bf j}_\|({\bf r}_\|,\,t)=0$, which
relates the diffusion current to the external electric field ${\bf
E}_\|({\bf r}_\|,\,t)$. In addition, by setting $T({\bf
r}_\|,\,t)=T_L$ as a constant, Equations\ (\ref{d4}) and (\ref{d14})
constitute the basic hydrodynamic equations for $\rho({\bf
r}_\|,\,t)$ and ${\bf j}_\|({\bf r}_\|,\,t)$. Although the particle
number conservation is enforced in this way, the energy of the
system is not conserved in general.
\medskip

The second-order moment of the Boltzmann equation is
formally written as

\[
\frac{2}{{\cal A}}\sum\limits_{{\bf k}_\|}\,\varepsilon_{k_\|}\,\frac{\partial
f({\bf r}_\|,\,{\bf k}_\|,\,t)}{\partial t}
\]
\[
=-\nabla_{{\bf
r}_\|}\cdot\left[\frac{2}{{\cal A}}\sum\limits_{{\bf
k}_\|}\,\varepsilon_{k_\|}\,{\bf v}_{k_\|}\,f({\bf r}_\|,\,{\bf
k}_\|,\,t)\right]-\frac{1}{\hbar}\left[\frac{2}{{\cal
A}}\sum\limits_{{\bf k}_\|}\,\varepsilon_{k_\|}\,{\bf F}_{k_\|}\cdot\nabla_{{\bf
k}_\|}f({\bf r}_\|,\,{\bf k}_\|,\,t)\right]
\]
\begin{equation}
+\frac{2}{{\cal A}}\sum\limits_{{\bf k}_\|}\,\varepsilon_{k_\|}\,{\cal W}^{\rm
(in)}_{{\bf k}_\|}\left[1-f({\bf r}_\|,\,{\bf k}_\|,\,t)\right]-\frac{2}{{\cal
A}}\sum\limits_{{\bf k}_\|}\,\varepsilon_{k_\|}\,{\cal W}^{\rm (out)}_{{\bf
k}_\|}\,f({\bf r}_\|,\,{\bf k}_\|,\,t)\ . \label{d15}
\end{equation}
The results in Eq.\,(\ref{d15}) can be simplified if we introduce
the energy-relaxation time $\tau_2$ through

\[
\frac{2}{{\cal A}}\sum\limits_{{\bf
k}_\|}\,\varepsilon_{k_\|}\,{\cal W}^{\rm (in)}_{{\bf k}_\|}\left[1-f({\bf
r}_\|,\,{\bf k}_\|,\,t)\right]-\frac{2}{{\cal A}}\sum\limits_{{\bf
k}_\|}\,\varepsilon_{k_\|}\,{\cal W}^{\rm (out)}_{{\bf k}_\|}\,f({\bf r}_\|,\,{\bf
k}_\|,\,t)
\]
\begin{equation}
=-\frac{\overline{\varepsilon}[T({\bf
r}_\|,\,t)]-\overline{\varepsilon}(T_L)}{\tau_2}\ , \label{d16}
\end{equation}
where $T_L$ is the lattice temperature, and $\tau_2$ can be evaluated using the calculated
non-equilibrium part of electron distribution $g^\prime_j(t)$ as well as
${\cal S}_{j,j'}(t)$ in Eq.\,(\ref{e7})
for fixed applied bias field, temperature and
device parameters. This leads Eq.\,(\ref{d15})
to the following electron power loss equation

\[
-\frac{\partial\overline{\varepsilon}[T({\bf
r}_\|,\,t)]}{\partial t}=\nabla_{{\bf r}_\|}\cdot{\bf S}_\|({\bf
r}_\|,\,t)+\frac{\overline{\varepsilon}[T({\bf
r}_\|,\,t)]-\overline{\varepsilon}(T_L)}{\tau_2}
\]
\begin{equation}
+e\left[{\bf E}_\|+\nu_F\left(\frac{{\bf
E}_\|\times{\bf B}_0}{E_\|}\right)\right]\cdot{\bf j}_\|({\bf
r}_\|,\,t)\ , \label{d17}
\end{equation}
where the second and last terms on the right-hand-side of this
equation corresponds to thermal energy exchange with lattice and
Joule heating, $\overline{\varepsilon}[T({\bf r}_\|,\,t)]=(2/{\cal
A})\sum\limits_{{\bf k}_\|}\varepsilon_{k_\|}\,f({\bf r}_\|,\,{\bf
k}_\|,\,t)$ is the average kinetic energy of electrons per area, and
${\bf S}_\|({\bf r}_\|,\,t)=(2/{\cal A})\sum\limits_{{\bf
k}_\|}\varepsilon_{k_\|}{\bf v}_{k_\|}\,f({\bf r}_\|,\,{\bf
k}_\|,\,t)$ is the electron surface energy flux per length. It is
easy to show that

\begin{equation}
\overline{\varepsilon}[T({\bf r}_\|,\,t)]\approx \frac{2}{{\cal
A}}\sum\limits_{{\bf
k}_\|}\varepsilon_{k_\|}\,f_0(\varepsilon_{k_\|},\,T,\,t)=N_c\left(k_{\rm
B}T\right)^3\,{\cal Q}_2(\eta)\ . \label{avg}
\end{equation}
By substituting Eq.\,(\ref{avg}) into Eq.\,(\ref{d17}), this lets us
find the spatial distribution of the electron temperature $T({\bf
r}_\|,\,t)$ at each time $t$.
\medskip

To simplify the third-order moment of the Boltzmann
equation, we still employ the momentum-relaxation time
approximation in Eq.\,(\ref{d6}). This leads to

\[
\frac{2}{{\cal A}}\sum\limits_{{\bf k}_\|}\,\varepsilon_{k_\|}\,{\bf
v}_{k_\|}\,f({\bf r}_\|,\,{\bf k}_\|,\,t)+\tau_1\,\frac{\partial}{\partial
t}\left[\frac{2}{{\cal A}}\sum\limits_{{\bf k}_\|}\,\varepsilon_{k_\|}\,{\bf
v}_{k_\|}\,f({\bf r}_\|,\,{\bf k}_\|,\,t)\right]
\]
\[
=-\tau_1\,\nabla_{{\bf
r}_\|}\cdot\left[\frac{2}{{\cal A}}\sum\limits_{{\bf
k}_\|}\,\varepsilon_{k_\|}\,{\bf v}_{k_\|}{\bf v}_{k_\|}\,f({\bf r}_\|,\,{\bf
k}_\|,\,t)\right]-\frac{\tau_1}{\hbar}\left[\frac{2}{{\cal A}}\sum\limits_{{\bf
k}_\|}\,\varepsilon_{k_\|}\,{\bf v}_{k_\|}\,{\bf F}_{k_\|}\cdot\nabla_{{\bf k}_\|}f({\bf
r}_\|,\,{\bf k}_\|,\,t)\right]
\]
\begin{equation}
+\frac{2}{{\cal A}}\sum\limits_{{\bf
k}_\|}\,\varepsilon_{k_\|}\,{\bf v}_{k_\|}\,f_0(\varepsilon_{k_\|},\,T,\,\mu_0)\ .
\label{d18}
\end{equation}
In addition, approximating $f({\bf r}_\|,\,{\bf k}_\|,\,t)$ on the
right-hand-side of Eq.\,(\ref{d18}) by
$f_0(\varepsilon_{k_\|},\,T,\,\mu_0)$, we get

\[
\frac{\partial {\bf S}_\|({\bf r}_\|,\,t)}{\partial
t}=-\frac{{\bf S}_\|({\bf r}_\|,\,t)}{\tau_1}
\]
\begin{equation}
-N_c\,\frac{\nu_F^2e}{2}\left\{2\left[{\bf
E}_\|+\nu_F\left(\frac{{\bf E}_\|\times{\bf
B}_0}{E_\|}\right)\right]\left(k_{\rm B}T\right)^2\,{\cal
Q}_1(\eta)+\frac{1}{e}\,\nabla_{{\bf r}_\|}\left[\left(k_{\rm
B}T\right)^3\,{\cal Q}_2(\eta)\right]\right\}\ . \label{d19}
\end{equation}
From Eq.\,(\ref{d19}), we are able to calculate the spatial
distribution of the electron energy flux ${\bf S}_\|({\bf
r}_\|,\,t)$ at each time $t$, from which the electron thermal conductivity can be obtained.
\medskip

Now let us summarize our findings in this subsection by presenting the complete set of equations describing the
moments of the Boltzmann equation.
Since Dirac electrons posses non-parabolic energy dispersion, the relevant moments are written as the powers of both energy and group velocity, i.e.

\begin{equation*}
\mathbf{M}_{n1,\,n2} (\mathbf{r}_{\parallel},\,t) = \frac{2}{\mathcal{A}}
\sum \limits_{\mathbf{k}_{\parallel}}\,(\mathbf{v}_{\mathbf{k}_{\parallel}})^{n1}(\varepsilon_{k_{\parallel}})^{n2}\,f(\mathbf{r}_{\parallel},\,
\mathbf{k}_{\parallel},\,t)\ ,
\end{equation*}
where $M_{0,\,0}$ has the meaning of electron density, $\mathbf{M}_{1,\,0}$ is the sheet current density, $M_{0,\,1}$ is
the average electron kinetic energy, and $\mathbf{M}_{1,\,1}$ is the electron surface energy flux.
The above moments are augmented with the electron temperature $T(\mathbf{r}_{\parallel},\,t)$.
Based on two calculated moments $M_{0,\,0}$ and $\mathbf{M}_{1,\,0}$, the momentum (or the group velocity) can be determined from the force-balance equation, i.e. $\mathbf{k}_{\parallel} =(\tau_1/\hbar)\,\mathbf{F}_{\mathbf{k}_{\parallel}}
(M_{0,\,0};\,\mathbf{M}_{1,\,0})$, where the momentum relaxation-time ($\tau_1$) approximation has been employed. In general,
the total force $ \mathbf{F}_{\mathbf{k}_{\parallel}}$, including the resistive force from electron scattering, must be found from the Maxwell equations. This point is further elucidated in the next paragraph.
Note that the simultaneous solution of the Boltzmann-Maxwell equations is known as the Vlasov-Maxwell equation\,\cite{vlasov}, which can be used to elevate the relaxation-time
approximation including the long-range Coulomb interaction. Our approach in this paper still retains the essence of the Maxwell equations but simplifies their introduction into the Boltzmann equation by means of the force-balance equation. As a result of this, it leaves us with the momentum and energy ($\tau_2$) relaxation times in the formalism.
This partial simplification of the Vlasov-Maxwell equations gives rise to the correct treatment of electron transport driven by a terahertz optical field.
\medskip

The system of equations for the moments is given by

\begin{gather}
\label{EQ:1}
\dot{M}_{0,\,0} = -\nabla_{\mathbf{r}_{\parallel}} \cdot \mathbf{M}_{1,\,0}\ ,\\
\label{EQ:2}
\dot{\mathbf{M}}_{1,\,0}=
-\frac{\mathbf{M}_{1,\,0}}{\tau_1}
-N_c\,\frac{\nu_F^2e}{2}
\left\{
{
\mathbf{F}_{\mathbf{k}_{\parallel}}
(M_{0,\,0};\,\mathbf{M}_{1,\,0})
\left(k_{\rm B}T\right)\,{\cal
Q}_0(\eta)+
\frac{1}{e}\,\nabla_{{\bf r}_\|}\left[\left(k_{\rm
B}T\right)^2\,{\cal Q}_1(\eta)\right]
}
\right\}\ ,\\
\label{EQ:3}
\dot{M}_{0,\,1} = - \nabla_{\mathbf{r}_{\parallel}} \cdot \mathbf{M}_{1,\,1}-
\frac{M_{0,\,1}-M_{0,\,1}(t \rightarrow -\infty)}{\tau_2} -
e \mathbf{F}_{\mathbf{k}_{\parallel}}
(M_{0,\,0};\,\mathbf{M}_{1,\,0}) \cdot \mathbf{M}_{1,\,0}\ ,\\
\label{EQ:4}
\dot{\mathbf{M}}_{1,\,1} = -\frac{\mathbf{M}_{1,\,1}}{\tau_1} -
\nu_F^2 e
\left\{
{
N_c\,\mathbf{F}_{\mathbf{k}_{\parallel}}
(M_{0,\,0};\,\mathbf{M}_{1,\,0}) \left({k_B T}\right)^2 {\cal Q}_{1} (\eta) + \frac{1}{2 e}
\nabla_{\mathbf{r}_{\parallel}} M_{0,\,1}
}
\right\}\ ,
\end{gather}
where $M_{0,\,1} = N_c (k_B T)^3{\cal Q}_{2} (\eta)$.
As mentioned before, these equations are intertwined with the Maxwell equations if one aims to look for the
self-consistent response of Dirac plasma to an incident optical field.
Another approach would be studying the ideal magneto-hydrodynamics of the plasma, which can be applied to the case
when magnetic lines are frozen into an electron plasma. Formally, this
corresponds to a vanishing total force ($\mathbf{F}_{\mathbf{k}_{\parallel}} = 0$) and to a decoupling of the
moments equations. Once those moments are obtained, it would provide the electromagnetic field inside the Dirac plasma via
Maxwell equations as described in the next subsection.

\subsection{Self-Consistent Field Equation}
\label{SEC:Model-2}

The Maxwell equations for the transverse magnetic component ${\bf
B}({\bf r},\,t)={\bf B}_\|({\bf r},\,t)+{\bf B}_\perp({\bf r},\,t)$
with ${\bf B}({\bf r}_\|,\,t)\equiv{\bf B}({\bf r}_\|,z=0,\,t)$, as
well as for the electric component ${\bf E}({\bf r},\,t)={\bf E}_\|({\bf
r},\,t)+{\bf E}_\perp({\bf r},\,t)$ with ${\bf E}({\bf
r}_\|,\,t)\equiv{\bf E}({\bf r}_\|,z=0,\,t)$, are given by

\begin{equation}
\nabla_{\bf r}\cdot{\bf B}({\bf r},\,t)=0\ ,\ \ \ \ \ \ \ \ \
\nabla_{\bf r}\cdot\left[\epsilon_r({\bf r})\,{\bf E}({\bf
r},\,t)\right]=0\ , \label{d23}
\end{equation}

\begin{equation}
\frac{\partial{\bf B}({\bf r},\,t)}{\partial t}=-\nabla_{\bf
r}\times{\bf E}({\bf r},\,t)\ ,\ \ \ \ \ \ \ \ \frac{\partial{\bf
E}({\bf r},\,t)}{\partial t}=\left[\frac{c^2}{\epsilon_r({\bf
r})}\right]\nabla_{\bf r}\times{\bf B}({\bf r},\,t)\ , \label{d22}
\end{equation}
where $\epsilon_r({\bf r})$ is the relative dielectric constant of
embedded host materials including the gate oxide material and induced
optical polarization field. The calculations of
these four equations can be performed by using the Delaunay-Voronoi
surface integration scheme.\,\cite{cell} The total
electromagnetic fields, ${\bf E}({\bf r},\,t)$ and ${\bf B}({\bf
r},\,t)$, are coupled to the moments of the Boltzmann
equation through the following boundary
conditions for $\{E_z,\,{\bf B}_\|\}$ at the two-dimensional
graphene sheet ($z=0$)

\[
\epsilon_r({\bf r}_\|,\,0^+)\,E_z({\bf
r}_\|,\,0^+,\,t)-\epsilon_r({\bf r}_\|,\,0^-)\,E_z({\bf
r}_\|,\,0^-,\,t)
\]
\begin{equation}
=\frac{e}{\epsilon_0}\left[n_s({\bf r}_\|,\,t)+N_{\rm
ion}-\rho({\bf r}_\|,\,t)\right]\ , \label{gauss}
\end{equation}

\begin{equation}
B_y({\bf r}_\|,\,0^+,\,t)-B_y({\bf
r}_\|,\,0^-,\,t)=-e\mu_0\,j_x({\bf r}_\|,\,t)\ ,
\end{equation}

\begin{equation}
B_x({\bf r}_\|,\,0^+,\,t)-B_x({\bf
r}_\|,\,0^-,\,t)=e\mu_0\,j_y({\bf r}_\|,\,t)\ ,
\end{equation}
where $N_{\rm ion}$ is the ion sheet density, and the total charge
neurility requires that

\begin{equation}
\frac{1}{{\cal A}}\,\int\,d^2{\bf r}_\|\,\rho({\bf r}_\|,\,t)=N_{\rm
ion}\ .
\end{equation}
Here, the Maxwell equations must be solved self-consistently with the Boltzmann moment equations in the previous subsection.\,\cite{last}
In addition, we have continuity conditions at the boundary $z=0$

\begin{equation}
E_{x,y}({\bf r}_\|,\,0^+,\,t)=E_{x,y}({\bf
r}_\|,\,0^-,\,t)\ ,
\end{equation}

\begin{equation}
H_{z}({\bf r}_\|,\,0^+,\,t)=H_{z}({\bf
r}_\|,\,0^-,\,t)\ .
\end{equation}
In Eq.\,(\ref{gauss}), $n_s({\bf r}_\|,\,t)$, which produces a
space-charge field, is the induced surface charge density by a gate
voltage $V_{\rm G}(t)$. For the graphene transistor structure, we also require that $\int\limits_{0}^{L_{\rm c}}
dx\,E_x({\bf r}_\|,\,t)=V_{\rm DS}(t)$ and $\int\limits_{0}^{L_{\rm
G}} dz\,E_z({\bf r},\,t)=V_{\rm G}(t)$, where we assume that the
conduction channel is in the $x$ direction with a channel length
$L_{\rm G}$ and $V_{\rm DS}(t)$ represents the applied source-drain
ac voltage. Also, $L_{\rm G}$ represents the gate electrode depth.

\section{Concluding Remarks}
\label{sec4}

In conclusion, we have found there is a minimum electron mobility for a graphene nanoribbon just before a threshold for
an applied electric field when entering into the nonlinear transport regime, which
is attributed to the gradual build-up of a frictional force from phonon scattering by the applied field. We
also predict a field-induced mobility enhancement right after this threshold value,
which is regarded as a consequence of initially-heated electrons in high energy states with a larger group velocity in an elastic-scattering
dominated graphene nanoribbon. Moreover, we have discovered that this mobility enhancement
reaches a maximum in the nonlinear transport regime as a combined result of an upper limit
for the carrier group velocity in a nanoribbon and a field-induced dramatically-increased phonon scattering in the system.
Finally, the threshold field can be pushed upward and the magnitude in the mobility enhancement
can be reduced simultaneously by a small correlation length for the line-edge roughness in
the high-field limit due to the occupation of high-energy states by field-induced electron heating.
\medskip

Additionally, we have formulated a self-consistent device simulation for
graphene transistors in the presence of field modulation within the
terahertz frequency range. This involves moment equations from the
Boltzmann equation up to the third order for fast carrier dynamics, as
well as full wave electromagnetics coupled to the Boltzmann equation
for describing both temporal and spatial dependence of the total field including the induced optical polarization field.
\medskip

When the electron density is increased in a graphene nanoribbon, multi-subband
transport occurs, the field-induced mobility enhancement is expected to be reduced,
and the effect of electron-electron scattering needs to be included. When the lattice
temperature becomes high, on the other hand, neither the optical phonon nor inter-valley
scattering should be neglected. As   the width of a nanoribbon is modulated, a
periodic potential along the ribbon will form, leading to a graphene nanoribbon super-lattice with additional mini-band gap opening at the Brillouin zone boundaries.
The results of the current research is expected to be very useful for our
 understanding and  design of high-power and high-speed graphene nanoribbon
 emitters and detectors in the terahertz frequency range.

\section*{Acknowledgement(s)}

This research was supported by contract \# FA 9453-07-C-0207 of AFRL. DH would like
to thank the Air Force Office of Scientific Research
(AFOSR) for its support. DH would also like to thank Prof. Xiang Zhang for hosting the Visiting
Scientist Program sponsored by AFOSR.

\newpage

\begin{references}
\bibitem{fn1}Such inter-valley scattering would require momentum transfer comparable with
the distance between $\mathbf{K}$ and $\mathbf{K}^\prime$ points.
\bibitem{fn2}A distinct line must be drawn between equilibrium distribution function
$f_j^{(0)}$ in absence of applied electric field and
stationary solution of the transport equation $\lim\limits_{t\rightarrow 0}\,f_j(t)$.
\end{references}
\end{document}